\documentclass[%
 reprint,
nofootinbib,
 amsmath,amssymb,
 aps,
]{revtex4-2}

\usepackage{graphicx}
\usepackage{dcolumn}
\usepackage{bm}
\usepackage{hyperref}
\usepackage[mathlines]{lineno}
\usepackage{diagbox}
\usepackage{tabularx}
\usepackage[dvipsnames]{xcolor}
\usepackage{adjustbox}

\begin{document}

\preprint{APS/123-QED}

\title{Share, Rotate, Split: The Effects of Group Work Role Distributions on Student Outcomes}
\author{Jacob Feinleib}
\affiliation{Laboratory of Atomic and Solid State Physics, Cornell University, Ithaca, New York 14853, USA}

\author{Matthew Dew}
\affiliation{Laboratory of Atomic and Solid State Physics, Cornell University, Ithaca, New York 14853, USA}

\author{N.G. Holmes}
\affiliation{Laboratory of Atomic and Solid State Physics, Cornell University, Ithaca, New York 14853, USA}

\date{\today}

\begin{abstract}
Education literature recommends many different strategies for structuring student group work in labs. Many of these strategies, however, have not been sufficiently evaluated for their effects on student outcomes. One prior study  
suggested that sharing roles, rather than splitting roles, in lab groups can boost students' physics interest and self-efficacy. Here, we expand upon this literature by evaluating the effects of a broader range of role distributions across several student outcomes from a large sample at two different institutions. We developed a survey item to probe the ways students distribute their roles in lab groups. 
The item asks for the percent of time in lab they spent working together on lab roles (sharing), working alone on roles but rotating each session (rotating), and working alone in the same role throughout the semester (splitting). We employed hierarchical linear modeling to measure the effects of these role distributions on student critical thinking, self-efficacy, perceived agency, belonging, and sense of recognition based on survey items specific to physics lab contexts. We found that role distributions did not differentially impact student critical thinking. We also found that sharing roles tended to have a positive impact on student attitudes; splitting had a negative effect on attitudes; and rotating fell in between. Statistical significance varied across these attitudinal outcomes. Our findings invite further research and controlled studies to better understand the apparent benefits of sharing, rotating, and splitting roles in introductory physics labs. 

\end{abstract}

\maketitle

\section{\label{sec:Intro}Introduction}
Group work is utilized in many physics lab classes, especially at the introductory level. Students typically work in groups of two to three \cite{geschwind_development_2024} to collect, analyze, and present data. While working in lab groups, students can perform a number of roles, for example, using equipment, taking notes, analyzing data, and managing the group \cite{quinn_group_2020}. Our prior work has investigated what roles students prefer to take on in lab \cite{holmes_evaluating_2022, dew_group_2024} and how students spend their class time on these roles \cite{quinn_group_2020, dew_group_2024, dew_structuring_2025}. Little research has been done, however, to understand how students distribute these roles in group work. 

One study sought to evaluate how different group work role distributions affect student outcomes. In an introductory physics lab, students reported a preference for a ``fair split'' between group members \cite{doucette_share_2022}. Students that reported being in groups where members ``participated equally'', however, were more likely to report that their interactions with their peers increased their physics interest and self-efficacy. The authors argue that although students might prefer to divide up work so that members specialize in a role (split), they benefit from each member working on the roles together (share); that is, when it comes to distribution of roles, ``share it, don't split it''.

In addition to groups sharing or splitting roles, students can also take turns working on each role (rotating). For example, with a group of three Students A, B, and C, these role distributions would look as follows:
\begin{itemize}
    \item Share: Students A, B, and C all focus on collecting data with equipment at the same time. While only one person at a time may be actively using the equipment, the other two students are actively discussing the data-collection and interacting with the student taking data.
    \item Rotate: Student A uses equipment, Student B analyzes data, and Student C takes notes. In a subsequent lab session, Student A analyzes data, Student B takes notes, and Student C uses equipment.
    \item Split: Throughout all their lab sessions together across the semester, Student A uses equipment, Student B analyzes data, and Student C takes notes.
\end{itemize}
Students may choose to use one or a combination of these role distributions across the semester. Unless otherwise instructed, students typically do not discuss how they distribute their roles \cite{quinn_group_2020} and report role distribution occurs ``naturally'' \cite{holmes_evaluating_2022}. When asked what they \textit{prefer}, one study found that students preferred to share roles \cite{holmes_evaluating_2022}; another found students similarly preferred to share \textit{or} split roles \cite{dew_group_2024}; while another indicated students preferred a ``fair split'', which could mean sharing, rotating, or splitting \cite{doucette_share_2022}. 

When groups opt to rotate or split roles, students work on tasks individually. The risk with splitting roles is that individual students may not gain experience with particular tasks. Some researchers recommend explicitly assigning students to roles and rotating the assignment, because it ensures they gain experience with multiple roles \cite{Rosser1998} and enables students to take on roles they may not otherwise \cite{heller_teaching_1992-1}. Other researchers argue against explicit role assignment, however, because it does not enable students to learn how to work as a group \cite{dew_group_2024} or because students tend to disregard role assignments~\cite{chang_when_2018}. We imagine an additional risk is that students may react negatively to being assigned a role they do not want or are not good at, potentially impacting their self-efficacy or sense of belonging.

While studies have argued for different recommendations for role distributions~\cite[e.g.,][]{johnson_educational_2009,johnson_making_1999,johnson_cooperative_1998,tanner_approaches_2003, tanner_structure_2013}, few base these recommendations on empirical results. Additionally, different forms of role distribution may be more or less useful for different student outcomes, whether that is learning, attitudes, or equity. For the purposes of student learning, there are theoretical reasons to expect certain role distributions are more productive.

Here, we leverage the ICAP (Interactive, Constructive, Active, and Passive) framework, which argues that students learn better at higher levels of engagement, where engagement spans four different levels \cite{chi_icap_2014}. From lowest to highest engagement, these are passive, active, constructive, and interactive. In the case of students learning about measurement uncertainty in experimental physics, the following are examples at each level:
\begin{itemize}
    \item Passive: A student \textit{receives} information from a lecture about the best ways to minimize uncertainty for an experimental set-up. 
    \item Active: A student \textit{manipulates} an experimental set-up following instructions on the best ways to minimize uncertainty. 
    \item Constructive: A student \textit{generates} ideas about different ways to minimize uncertainty and tests them on an experimental set-up.
    \item Interactive: A student \textit{dialogues} with another student to generate and test ways to reduce uncertainty for an experimental set-up.
\end{itemize}


For learning to think critically about experiments, therefore, we expect that the dialogue about the experiments (interactive) is more important for student learning than simply carrying out the tasks (active). For example, discussing the best ways to set up the equipment and collect data in order to minimize uncertainties should be more important for students' critical thinking than carrying out the resulting procedures. Thus, we may expect that students sharing roles most effectively motivates constructive dialogue, as students need to make decisions together about how to set up the equipment and collect data, for example. Rotating roles may next most effectively motivate constructive dialogue because, while each student is responsible for a role in a given week, other students have common proficiency with the role to support constructive dialogue. Splitting roles may least effectively motivate constructive dialogue, as each student has a unique specialty that they bring to the discussion, meaning students may be more likely to take the recommendations of the ``expert'', rather than constructively dialogue.

We see no clear ordering, however, of group work role distributions most productive for student attitudes. Each role distribution likely provides unique benefits dependent upon the attitude measure. For example, students who split roles may develop a sense of expertise (or self-efficacy), while students who rotate roles may feel capable of doing each individual role. Alternatively, students who share roles, may feel more socially integrated (develop a sense of belonging or recognition) because they have to interact with their team members throughout more stages of the lab. 




Prior work suggests that sharing roles is more productive than splitting when it comes to students' physics interest and self-efficacy development \cite{doucette_share_2022}.
The study, however, has several limitations that motivate further investigation. First, rather than measuring students' interest and self-efficacy at the start and end of the semester, the survey items asked students if their interactions with peers increased their interest and self-efficacy. 
Second, students were asked only a binary question about their role distribution -- whether or not both they and their partner ``participated equally,'' which the authors interpreted as sharing roles. It is possible that students split or rotated roles in a way they believed was equal participation; the study did not comment on students' interpretations from think-aloud interviews. Finally, the binary question does not elucidate differences between rotating and splitting roles, which prior work has noted students perceived differently \cite{holmes_evaluating_2022, dew_group_2024}. 

Given the compelling claim, ``share it, don't split it,'' this study invites further research. While the study investigated whether students found peer interactions positively affected their physics interest and self-efficacy, it remains an open question whether their learning or other constructs were impacted by their group work role distribution. Importantly, do student attitudes -- rather than students' perceptions of how \textit{peer interactions' affected their} attitudes -- improve? Additionally, the study looked at ``conceptual guided-inquiry'' labs at a single institution. Here we investigate a different context: skills-based, open-inquiry labs at two different institutions. Altogether, we seek to expand the prior investigation by evaluating more student outcome constructs across a larger population and additional instructional contexts. 

In this study, we test Doucette et al.'s claim by evaluating the effect of different group work role distributions on a range of student outcomes. Specifically, our research questions are as follows:
\begin{enumerate}
    \item How do students report distributing their group work roles in introductory physics labs? 
    \item Which group work role distributions lead to better student outcomes in open-inquiry physics labs?
\end{enumerate}

\section{\label{sec:Methods}Methods}
We detail our methodology in this section. First, in Subsection \ref{subsec:CourseContext}, we discuss our course context and student population. In Subsection \ref{subsec:OutcomeConstructs}, we detail the measures we use for student outcomes. In Subsection \ref{subsec:SurveyDevelopment}, we discuss the survey items we developed to quantify group work role distributions. In Subsection \ref{subsec:MultipleImputation}, we explain how we used multiple imputation to account for missing data. Finally, in Subsection \ref{subsec:HLMAnalysis}, we share our methodology for addressing our research questions via hierarchical linear modeling. 

\subsection{\label{subsec:CourseContext}Course Context and Participants}
Our study focuses on three introductory physics lab courses: two at Institution A, a public university in the Southern US, and one at Institution B, a private university in the Northeast US. These three lab courses use the Structured Quantitative Inquiry framework \cite{Smith2020}. They are considered ``open-inquiry labs''~\cite{buck_research_2008}: in most lab activities, students are provided with a problem or question and some theory/background, but have the opportunity to design their own procedures, choose how to analyze their data, and draw conclusions supported by their data. These labs also focus on developing experimentation skills, such as designing experiments and analyzing data,  and attitudes, such as epistemological understandings of measurement (see Ref.~\cite{holmes_operationalizing_2019} for a description of the learning goals). We collected survey data across two semesters. The two courses at Institution A (Courses A1 and A2) each had the same instructors for both semesters, whereas the course at Institution B (Course B) had a different instructor each semester.

Course A1 covered mechanics topics while Course A2 covered E\&M, optics, and modern physics topics. Both labs were one credit hour and were corequisites to the introductory physics lecture courses. Students enrolled in these labs were in one of three lecture tracks: algebra-based physics, calculus-based physics for life science majors, and calculus-based physics for engineering majors. The labs were typically taken earlier in student degree sequences by engineering majors and later by life science majors and those enrolled in the algebra-based sequence. See Table \ref{tab:demographics} for more detailed course demographics. Course A1 concluded with a lab practical exam while Course A2 concluded with a project-based lab where groups created and tested their own research questions (a level of authentic inquiry, based on Ref.~\cite{buck_research_2008}).

Course B was a one credit-hour lab course covering mechanics and E\&M topics and concluded with a project-based lab similar to Course A2. This course was primarily taken by freshman and sophomores in physics and engineering. 

In Courses A1 and A2, students had a three hour lab session in which to collect data and had to submit their lab notes and analysis by the start of the following lab session. In Course B, students completed their labs entirely within the two-hour lab session.

All three courses involved extensive group work and a majority of each student's final grade was determined by group assignments. Additionally, working collaboratively was an explicit learning objective in each course. In all three courses, students self-selected into groups of three. In situations where a group of three could not be made, students in Courses A1 and A2 formed groups of two while students in Course B formed groups of four. Students changed lab groups twice throughout the semester, every three to four lab sessions.

\begin{table*}[]
\caption{Course-level demographics by semester and course collected through the pre- and post-surveys. If a student's demographic response differed on the post-survey from the pre-survey, we used the post-survey answer.}
\label{tab:demographics}
\begin{ruledtabular}
\begin{tabular}{lcccccc}
 & \multicolumn{3}{c}{Semester 1} & \multicolumn{3}{c}{Semester 2} \\ \cline{2-4} \cline{5-7}
Student-level variables & A1 & A2 & B & A1 & A2 & B\\ \hline
All & 815 & 665 & 618 & 864 & 560 & 380\\
Gender &  &  &  &  &  & \\
\quad Woman & 462 & 375 & 284 & 543 & 308 & 164\\
\quad Man & 336 & 268 & 316 & 295 & 230 & 196\\
\quad Non-binary or other & 13 & 9 & 9 & 13 & 12 & 9\\
\quad Did not disclose & 4 & 13 & 9 & 13 & 10 & 11\\
Race or ethnicity &  &  &  &  &  & \\
\quad American Indian or Alaska Native & 7 & 5 & 3 & 10 & 5 & 1\\
\quad Asian & 277 & 257 & 275 & 346 & 215 & 239\\
\quad Black or African American & 48 & 35 & 64 & 54 & 36 & 13\\
\quad Hispanic or Latino & 244 & 169 & 72 & 244 & 146 & 22\\
\quad Middle Eastern or North African & 31 & 20 & 15 & 20 & 12 & 9\\
\quad Native Hawaiian or Other Pacific Islander & 3 & 2 & 3 & 6 & 0 & 1\\
\quad White & 282 & 230 & 221 & 255 & 198 & 113\\
\quad Some other race or ethnicity & 1 & 2 & 10 & 4 & 0 & 1\\
\quad Did not disclose & 4 & 13 & 9 & 13 & 10 & 11\\
Class standing &  &  &  &  &  & \\
\quad Freshman & 396 & 52 & 590 & 49 & 8 & 301\\
\quad Sophomore & 278 & 101 & 20 & 189 & 162 & 70\\
\quad Junior & 106 & 402 & 5 & 531 & 255 & 7\\
\quad Senior & 31 & 102 & 3 & 86 & 130 & 2\\
\quad Other & 0 & 4 & 0 & 3 & 2 & 0\\
\quad Did not disclose & 4 & 13 & 9 & 13 & 10 & 11\\
Major &  &  &  &  &  & \\
\quad Physics, astronomy, or engineering physics & 105 & 25 & 39 & 32 & 42 & 72\\
\quad Engineering & 259 & 82 & 488 & 49 & 135 & 246\\
\quad Life science or biology & 276 & 394 & 9 & 520 & 259 & 10\\
\quad Other physical science & 33 & 41 & 44 & 42 & 31 & 39\\
\quad Other & 0 & 4 & 0 & 3 & 2 & 0\\
\quad Did not disclose & 4 & 13 & 9 & 13 & 10 & 11\\
Parent's highest level of education &  &  &  &  &  & \\
\quad Did not complete high school & 47 & 25 & 23 & 39 & 20 & 14\\
\quad High school/GED & 139 & 97 & 103 & 132 & 80 & 48\\
\quad Some college (but did not complete college) & 64 & 45 & 22 & 58 & 43 & 10\\
\quad Associate's degree (2 year degree) & 22 & 34 & 10 & 32 & 10 & 8\\
\quad Bachelor's degree & 221 & 139 & 96 & 227 & 149 & 59\\
\quad Master's degree & 180 & 150 & 191 & 216 & 148 & 106\\
\quad Advanced graduate degree & 118 & 140 & 138 & 125 & 89 & 112\\
\quad Not sure & 3 & 8 & 12 & 9 & 4 & 5\\
\quad Prefer not to disclose & 21 & 27 & 23 & 26 & 17 & 18\\
\end{tabular}
\end{ruledtabular}
\end{table*}

We gathered data through surveys sent to students at the beginning and end of each semester. Students were incentivized to complete the surveys with a small fraction of their grade being determined by completion of the survey (independent of student consent to participate in research). The survey was assigned to be completed outside of class. 
Response rates by course are in Table \ref{tab:record-counts}. We used multiple imputation for missing data, detailed in Section \ref{subsec:MultipleImputation}.

\begin{table*}[]
\caption{Number of participants by semester and course. To be counted as ``finished'', students had to meet the following criteria: the student completed the survey; the student consented to  participate in research; the student is at least 18 years of age; the student spent at least 30 seconds on any page; and the student provided some sort of identification (i.e. name, student ID). Data was imputed according to Subsection \ref{subsec:MultipleImputation}. }
\label{tab:record-counts}
\begin{ruledtabular}
\begin{tabular}{lcccccc}
 & \multicolumn{3}{c}{Semester 1} & \multicolumn{3}{c}{Semester 2} \\ \cline{2-4} \cline{5-7}
 & A1 & A2 & B & A1 & A2 & B\\ \hline
Finished pre-survey & 760 & 606 & 577 & 773 & 475 & 362\\
Finished post-survey & 371 & 374 & 469 & 563 & 323 & 283\\
Finished both & 316 & 315 & 428 & 472 & 238 & 265\\ \hline
Imputed pre-survey & 55 & 59 & 41 & 91 & 85 & 18\\
Imputed post-survey & 444 & 291 & 149 & 301 & 237 & 97\\ \hline
Total after imputation & 815 & 665 & 618 & 864 & 560 & 380\\
\end{tabular}
\end{ruledtabular}
\end{table*}

\subsection{\label{subsec:OutcomeConstructs}Outcome Constructs}
We measured student outcomes by testing critical thinking skills and student attitudes towards experimental physics, namely self-efficacy, perceived agency, belonging, and recognition. 

\subsubsection{Development of the Outcome Constructs}

Critical thinking was measured using the Physics Lab Inventory of Critical thinking (PLIC), which has been previously evaluated for reliability and validity~\cite{walsh_quantifying_2019}. An additional study~\cite{heim_what_2022} motivated us to remove two items from the PLIC related to evaluating methods. The study found that the two items did not sufficiently engage students' critical thinking as compared to the final evaluating methods item. The study also found that removing those items did not affect students' reasoning on the final evaluating methods item.

The development of the attitudes items was partially reported in Refs.~\cite{kalender_restructuring_2021, kalender_sense_2020}. Items were either taken from or developed based on existing survey instruments or literature about each construct. Two self-efficacy items came from Refs.~\cite{kost-smith_characterizing_2011, kalender_gendered_2019}. The other self-efficacy items were inspired by the task-specific items in Ref.~\cite{kost-smith_characterizing_2011}, but developed to relate to specific tasks students carry out in introductory physics labs. Three of the perceived agency items came from Ref.~\cite{tapal_sense_2017}, again adapted to be in the context of physics labs. One perceived agency item was created to reflect Scardamalia and Bereiter's~\cite{scardamalia_knowledge_2014} definition of epistemic agency. These four items were previously used in Ref.~\cite{kalender_restructuring_2021} to study students' sense of agency. All four belonging items came from Refs.~\cite{kalender_gendered_2019, good_why_2012}, though contexts were adapted from ``in a math setting'' and ``in this class'' to being ``in a physics lab.'' Finally, all four recognition items were adapted from those in Ref.~\cite{lock_impact_2019}, though recognition from parents/relatives/friends/physics teacher were adapted to be from lab instructors/lab peers and physics instructors/physics peers outside of lab. 

Researchers conducted interviews with physics and engineering majors enrolled in introductory physics lab courses at Institutions A and B to evaluate the construct validity~\cite{adams_development_2011}. During the interviews, students were asked to complete the survey and think out loud while they responded to the items. They were also asked to clarify their thinking and how they were interpreting items. The interviews led to some items being reworded and some items being dropped from the survey. Interviews continued until saturation was achieved (15 total interviews). We further checked the validity of these outcome construct items using confirmatory factor analysis, as described below.\footnote{Two belonging items were excluded from the analysis early on in the confirmatory factor analysis due to low factor loadings. The numbers we report below are from the final confirmatory factor analysis without these items.}

\subsubsection{Confirmatory Factor Analysis Assumptions}
We used confirmatory factor analysis to evaluate the extent to which the survey items reflect the theoretical relationships between constructs. First, we evaluated whether the data meet the assumptions necessary for confirmatory factor analysis,following recommendations from Knetka et al. \cite{knekta_one_2019}. 

Combining the pre- and post-semester survey responses, we had 5904 responses.\footnote{This response count is less than the sum of the first two rows from Table \ref{tab:record-counts}. There are 32 fewer responses than that sum because these students did not respond to all four construct-related questions for that survey response.} Of these 5904 responses, we were missing at least one construct-related item for 2.9\% of responses. We investigated our data for outliers and found that student responses spanned the full Likert scale for all items, although some had narrow distributions. We calculated the Mahalanobis distances \cite{raykov2008introduction, tabachnick2007using} for our data to identify potential outliers. Upon inspection, we did not find any reasons (such as students selecting the same choice for all questions) to drop these potential outliers. 

To evaluate factorability, we used Kaiser's measure of sampling adequacy. We obtained a measure of sample adequacy of 0.93, well above the threshold of 0.6 \cite{knekta_one_2019}. We also checked the inter-item correlations and found that the minimum correlation among items from the same factor was 0.43, which is above the 0.30 threshold \cite{knekta_one_2019}.

Next, we checked the normality and linearity of our data. We assessed univariate normality with skewness and kurtosis. No item should have a skewness or kurtosis above $|2.0|$ \cite{knekta_one_2019}; our highest value among either of these was 1.16. We checked multivariate normality using Mardia's multivariate normality test and found statistically significant skewness ($p<0.001$) and kurtosis ($p<0.001$) for our data. We addressed this in our analysis by using an estimator for confirmatory factor analysis that is robust against non-normality. We checked linearity within factors by plotting--for each factor--student scores for every item against their scores on that factor's other items. We also calculated Spearman's Rank Order Correlation Coefficient between factor items and found all coefficients were greater than 0.4 and statistically significant ($<0.001$), indicating sufficient linearity \cite{knekta_one_2019}. 

Lastly, we checked multicollinearity using variance inflation factors. Our largest value was 4.03, which is less than the maximum threshold of 10 \cite{knekta_one_2019} indicating that our data are not multicollinear.

With all these checks, our data meet the assumptions necessary for confirmatory factor analysis, so long as we use a methodology robust to non-normality. The full set of figures and values used for checking our assumptions are available in the Supplemental Material \cite{supplementalMaterial}.

\subsubsection{Confirmatory Factor Analysis Results}

We performed confirmatory factor analysis with a four-factor structure: one factor each for self-efficacy, perceived agency, belonging, and recognition (Table~\ref{tab:attitudes}). We used maximum likelihood estimation with robust standard errors to account for the non-normality of our data. The chi-squared test of model fit was statistically significant ($\chi^2 = 2963.179, df = 203, p < 0.001$), suggesting our data do not the fit the model. The chi-squared value typically increases as the data size increases, however, which may mean our model is being unfairly rejected due to small differences \cite{knekta_one_2019, kline2023principles}. We consider this limitation in the context of other fit indices.  

We obtained a Comparative Fit Index (CFI) of 0.956, a Root Mean Square Error of Approximation (RMSEA) of 0.056, and a Standardized Root Mean Square Residual (SRMR) of 0.030. All of these measures of goodness-of-fit meet the desired thresholds; CFI above 0.95, RMSEA below 0.06, and SRMR below 0.08 \cite{knekta_one_2019}. 

\begin{figure}
    \centering
    \includegraphics[width=1\linewidth]{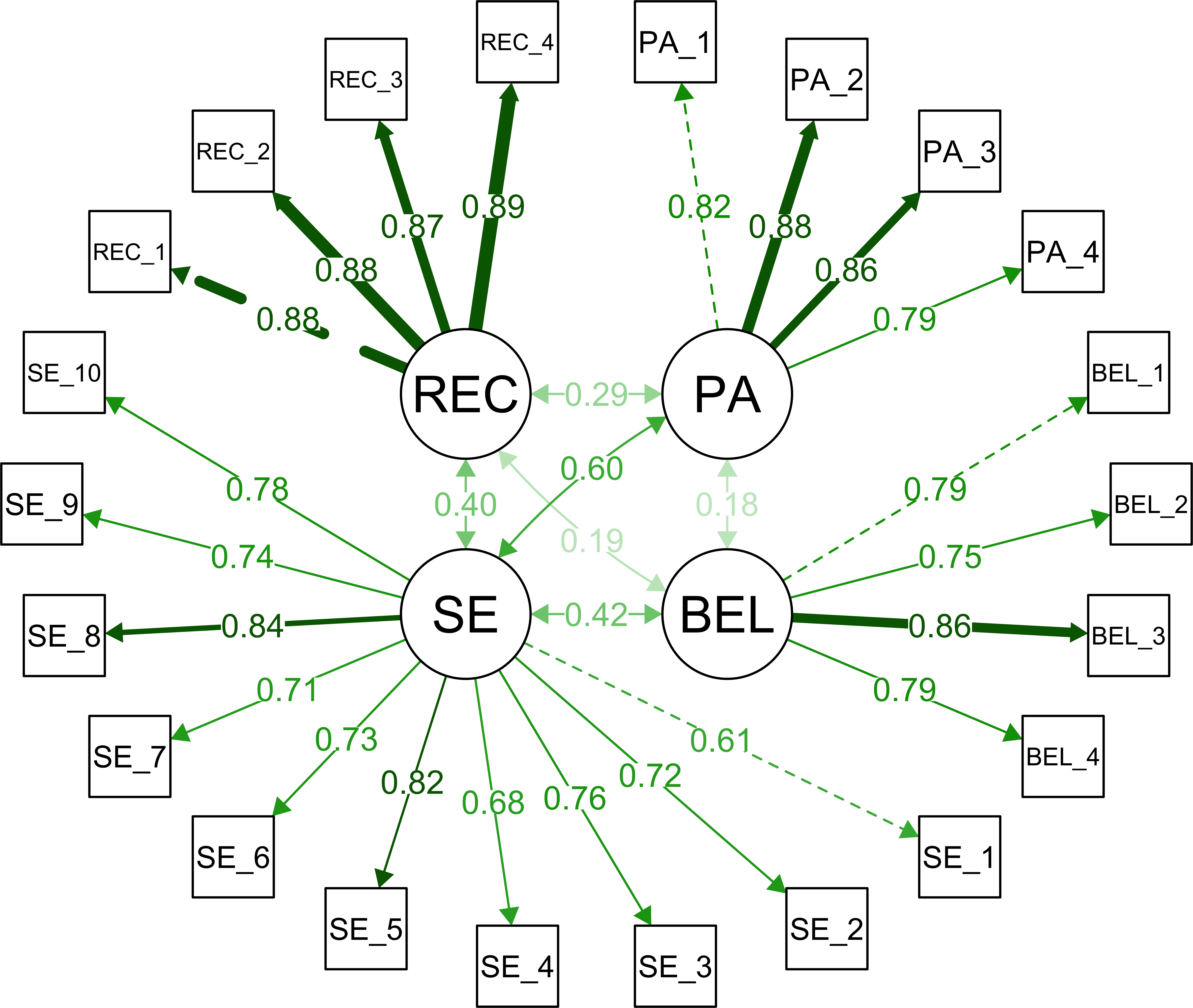}
    \caption{Factor loadings from confirmatory factor analysis of outcome constructs. "SE" is self-efficacy, "PA" is perceived agency, "BEL" is belonging, and "REC" is recognition. Anchor items for each factor are denoted with a dashed line. Line thickness and shade denote factor loading magnitude. Table \ref{tab:attitudes} gives the full text for each survey item.}
    \label{fig:CFA}
\end{figure}

Our factor loadings are given in Figure \ref{fig:CFA} and Table \ref{tab:attitudes} (the latter with the full list of questions). All items, except two, had factor loadings above 0.7. The two items that did not meet this threshold were SE\textunderscore1 (``Express my opinions when others disagree with me during the lab,'' which had a loading of 0.61) and SE\textunderscore4 (``Interpret graphs of my measurements,'' which had a loading of 0.68). We kept these items for several reasons: the loadings are both near the threshold, SE\textunderscore1 was almost identical to an item from Ref.~\cite{kost-smith_characterizing_2011}, and both items were constructed as important items to the construct of lab self-efficacy. 

\begin{table*}
\caption{Survey items by construct. The self-efficacy items prompt reads, ``Please rate how confident you are that you can do each of the following things in a physics lab course.''  The answer options for the self-efficacy items ranged from ``Not Confident" to ``Confident" on a five point Likert scale. The other item prompts read, ``Please indicate how well you agree with the following statements'' and had answer options on a five point Likert scale ranging from ``Strongly Disagree" to ``Strongly Agree." Items marked with * were reverse coded.\label{tab:attitudes}}
\begin{ruledtabular}
\begin{tabular}{lllc}
  Construct & Item & Item & Factor \\
  & number & & loading \\ \hline
  Self-efficacy & & & \\
  & SE\textunderscore1 & Express my opinions when others disagree with me during the lab. & 0.61 \\
  & SE\textunderscore2 & Overcome any problems I encounter in an experiment. & 0.72 \\
  & SE\textunderscore3 & Interpret experimental outcomes taking into account experimental uncertainty. & 0.76 \\
  & SE\textunderscore4 & Interpret graphs of my measurements. & 0.68 \\
  & SE\textunderscore5 & Design reliable experiments with the available equipment. & 0.82 \\
  & SE\textunderscore6 & Make accurate predictions about experimental outcomes. & 0.73 \\
  & SE\textunderscore7 & Comfortably take charge of the equipment in the physics lab. & 0.71 \\
  & SE\textunderscore8 & Design an experiment that answers my research question. & 0.84 \\
  & SE\textunderscore9 & Generate further research questions based on my observations in the lab. & 0.74 \\
  & SE\textunderscore10 & Design an experiment in a physics lab where I can find something I did not know before. & 0.78 \\
  Perceived & & & \\
  agency & & & \\
  & PA\textunderscore1 & I am in control of setting the goals for the experiments. & 0.82 \\
  & PA\textunderscore2 & I have the freedom to design and conduct the best possible experiment to attain my goals. & 0.88 \\
  & PA\textunderscore3 & I am in control of choosing the appropriate analysis tools to evaluate experimental outcomes. & 0.86 \\
  & PA\textunderscore4 & I am in control of doing interesting experiments in a physics lab. & 0.79 \\
  Belonging & & & \\
  & BEL\textunderscore1 & I feel like an outsider in a physics lab.* & 0.79 \\
  & BEL\textunderscore2 & When I get a poor grade on an experiment, I feel that maybe I don’t belong in a physics lab.* & 0.75 \\
  & BEL\textunderscore3 & Sometimes I worry that I do not belong in a physics lab.* & 0.86 \\
  & BEL\textunderscore4 & When I am in a physics lab, I wish I could fade into the background and not be noticed.* & 0.79 \\
  Recognition & & & \\
  & REC\textunderscore1 & My lab peers see me as a physics person. & 0.88 \\
  & REC\textunderscore2 & My lab instructors see me as a physics person. & 0.88 \\
  & REC\textunderscore3 & My physics peers outside of lab see me as a physics person. & 0.87 \\
  & REC\textunderscore4 & My physics instructors outside of lab see me as a physics person. & 0.89 \\
\end{tabular}
\end{ruledtabular}
\end{table*}

\subsection{\label{subsec:SurveyDevelopment}Roles Distribution Survey Development}
We iteratively developed a survey item to measure the ways in which students distributed their roles during their collaborative lab work. The goal of the item was to serve as a post-semester estimate of how students tended to distribute their roles across the full course. First, we distributed a pilot open response question on a post-survey to an introductory physics lab course at Institution B. The open response question read:
    \begin{quotation}
        \noindent Use the space below to explain how your group split up the above roles. (i.e. did one person do a role, did multiple people work independently, or did everyone work together?)
    \end{quotation}
We first read each text response ($N=1037$) to this question and found a number of themes common across responses:
\begin{itemize}
    \item Working on all roles together,
    \item Individuals taking on roles that they did not take on during the previous lab session, and
    \item Individuals specializing in and staying with roles/doing roles they had experience with.
\end{itemize}
These themes expand upon the share/split framework~\cite{doucette_share_2022} to include the behavior of students working individually but switching roles between sessions, which we refer to as rotate. The themes also echo those found in an interview study conducted at Institution B~\cite{holmes_evaluating_2022}.

We further inspected responses that included a specific reason indicating why a group distributed roles in a certain way ($N=80$). 
After excluding themes that appeared with low frequency, we identified five recurring themes in student responses: 
\begin{itemize}
    \item Sharing so everyone could gain experience
    \item Rotating so everyone could experience different roles
    \item Splitting so people could do what they enjoyed
    \item Splitting because of logistical constraints
    \item Naturally falling into roles
\end{itemize}
Students generally used the word ``naturally'' to mean there was no rationale motivating their distribution of lab roles.

We converted the open response question into a closed response format where students indicated how and why they distributed their group work roles. The answer choices for this question were the three common types of role distribution from the open response question: sharing, rotating, and splitting. Each choice also included a positive reason to choose it so that no option seemed undesirable. For example, one choice read, ``We split roles so people could do what they enjoyed.” Here, enjoyment was the positive reason to choose splitting. The question thus read: 

\begin{quotation}
   \noindent During your time \textit{in the lab}, how did your group distribute the lab roles (equipment, notes, analysis)? Why?
   \begin{itemize}
    \item We all worked together on every role so that everyone could learn about and experience every role.
    \item We divided the roles and rotated who worked on each role so that everyone could learn about and experience every role.
    \item We divided the roles and kept the same roles each week so each person could do what they enjoyed or were experienced with.
    \item We divided the roles based on logistical constraints (i.e. data or code was on one person’s computer).
    \item We naturally fell into roles without any discussion.
    \item Other (with an option to include a text response)
\end{itemize}
\end{quotation}

In transferring this question to an online survey format, we used a constant-sum response, where students could allocate up to 100\% of their time among the options. An allocation of 100\% to an item would indicate that their group(s) distributed their lab roles in that way throughout the whole semester. An allocation of 50\% to an item would indicate that their group(s) distributed their lab roles in that way about half of the time.

We conducted interviews with four students in Course B to evaluate the question and answer choices. In these interviews, interviewees answered the questions about their most recent lab course. As interviewees completed the survey, we asked them to think out loud, particularly to articulate how they interpreted each question and which specific lab roles they had in mind for each question. We had three main takeaways from this set of interviews. First, the ``natural'' option was overly broad and could encompass both sharing and splitting roles. With the constant sum format, we wanted to avoid redundancy. Second, we did not need two separate options for splitting roles with different reasons. Our research questions were more concerned with the actual role distribution than the student's rationale. Third, we found that students struggled to distinguish working together cognitively (such as on decision-making tasks) with dividing or sharing the hands-on roles. 

These findings led us to reduce the number of options from six to four and we reworded the options to have the positive reason be in parentheses to act as a suggestion rather than a criteria. In addition, we adjusted the question to specify that we were asking about hands-on roles as opposed to decision-making roles. The prompt and answer choices read:
\begin{quotation}
\noindent During your time in the lab, how frequently did your group organize \textbf{physical, hands-on lab roles} (equipment, notes, and analysis) in the following ways?
Enter a percentage of time for each option below where 0\% is ``Never'' and 100\% is ``Always''. Your percentages should add up to 100\%.
\begin{itemize}
    \item We worked together on roles (for example, so that we could bounce ideas off each other for each role)
    \item We divided the roles and rotated who worked on each role (for example, so that everyone could learn about and experience every role)
    \item We divided the roles and kept the same roles each week (for example, so that each person could do what they enjoyed or were experienced with)
    \item Other
\end{itemize}
\end{quotation}
Students that selected ``Other'' were prompted with a text entry box to describe in more detail.

We then interviewed six students taking Course A1 or A2, with the goal of validating the survey questions with a new population of students. The interviewees generally interpreted the items as intended. The only adjustment made from these interviews was to bold the words ``rotated'' and ``same'' in the second and third answer choices, respectively. This change occurred after the second interview. The following four interviews brought us to saturation.

\subsection{\label{subsec:MultipleImputation}Multiple Imputation}

Our outcome constructs were measured through pre- and post-semester surveys and the group work role distribution item was implemented only on the post-semester survey. Matching across pre- and post-surveys led to missing data and so we follow prior recommendations to impute data \cite{nissen_missing_2019}. We imputed our data following directions from \cite{Woods2024}, particularly their supplemental decision tree \cite{Woods2021}.

Our percentages of missing data for each variable in each course by semester are given in Table \ref{tab:missingness-rates}. While these missingness percentages vary across courses, prior work has found that multiple imputation is robust across various situations, such as regression models with up to 18 predictors, missingness percentages up to 50\%, and sample sizes as small as $N=50$ \cite{graham1999performance}. 

\begin{table}[]
\caption{Percentage of missingness for survey items on the pre- and post-surveys by course and semester.}
\label{tab:missingness-rates}
\begin{ruledtabular}
\begin{tabular}{lcccccc}
 & \multicolumn{3}{c}{Semester 1} & \multicolumn{3}{c}{Semester 2} \\ \cline{2-4} \cline{5-7}
Survey item & A1 & A2 & B & A1 & A2 & B\\ \hline
Pre-survey & & & & & & \\
\quad PLIC & 6.7\% & 8.9\% & 6.6\% & 10.5\% & 15.2\% & 4.7\%\\
\quad Self-efficacy & 9.2\% & 10.7\% & 8.7\% & 12.4\% & 17.0\% & 5.0\%\\
\quad Perceived & 8.2\% & 9.8\% & 9.1\% & 12.6\% & 17.3\% & 5.0\%\\
\quad agency & & & & & & \\
\quad Belonging & 7.7\% & 10.1\% & 8.4\% & 12.5\% & 16.4\% & 5.3\%\\
\quad Recognition & 7.9\% & 10.2\% & 7.8\% & 12.2\% & 17.0\% & 5.0\%\\
Post-survey & & & & & & \\
\quad PLIC & 54.5\% & 43.8\% & 24.1\% & 34.8\% & 42.3\% & 25.5\%\\
\quad Self-efficacy & 55.2\% & 44.4\% & 25.1\% & 35.8\% & 43.2\% & 26.1\%\\
\quad Perceived & 55.0\% & 44.2\% & 24.3\% & 35.3\% & 43.2\% & 25.5\%\\
\quad agency & & & & & & \\
\quad Belonging & 55.2\% & 44.2\% & 24.6\% & 35.5\% & 43.0\% & 25.5\%\\
\quad Recognition & 55.2\% & 44.4\% & 24.6\% & 35.3\% & 43.2\% & 25.5\%\\
\quad Share & 54.5\% & 43.8\% & 24.1\% & 34.8\% & 42.3\% & 25.5\%\\
\quad Rotate & 54.5\% & 43.8\% & 24.1\% & 34.8\% & 42.3\% & 25.5\%\\
\quad Split & 54.5\% & 43.8\% & 24.1\% & 34.8\% & 42.3\% & 25.5\%\\
\end{tabular}
\end{ruledtabular}
\end{table}

First, before imputing data, we evaluated whether we had access to other variables related to our missingness. For missing post-semester data, the most common explanatory variable used in imputation is final grade \cite{nissen_missing_2019}. Students with lower final grades are less likely to complete post-semester surveys \cite{nissen_missing_2019}. We only had access to grade data for Semesters 1 and 2 of Course B. We did not have access to grade data for Courses A1 and A2 for either semester. Thus, we checked if post-semester survey missingness was explained by pre-semester survey data for Courses A1 and A2. 

For Courses A1 and A2, we checked if pre-semester PLIC scores were a statistically significant predictor for post-semester survey missingness. Using a t-test, we found pre-semester PLIC scores were a statistically significant predictor for both Course A1 ($p=0.026$) and Course A2 ($p<0.001$). This test indicated that students with higher pre-semester PLIC scores were more likely to complete the post-semester survey. We also ran a logistic regression for each course controlling for random effects due to the semester offering and found that pre-semester PLIC scores were still predictors of post-semester survey missingness for Course A1 ($\beta = -0.835, p = 0.048$) and Course A2 ($\beta = -2.492, p < 0.001$). 

In Course B, we inspected whether final grades were a statistically significant predictor for post-semester survey missingness. For final grades, we used final letter grades on a 4.3 scale, which is a typical 4.0 scale with +'s adding 0.3 and -'s subtracting 0.3 from the typical letter grade point. We ran a $t$-test across all Course B data and found students with higher grades were more likely to complete the survey ($p < 0.001$). We similarly ran a logistic regression controlling for random effects due to the semester offering and found that grade was still a predictor of post-semester survey missingness ($\beta = -4.283, p<0.001$). 


For all three courses, we checked if post-semester PLIC scores predicted pre-semester survey missingness. Using t-tests, we found that post-semester PLIC scores predicted pre-survey missingness for Course A1 ($p=0.016$), Course A2 ($p=0.003$), and Course B ($p=0.002$). We also ran logistic regressions for each course controlling for random effects due to the semester offering and found post-semester PLIC scores were still a predictor for pre-semester survey missingness for Course A1 ($\beta = -1.718, p=0.013$), Course A2 ($\beta = -2.299, p=0.001$), and Course B ($\beta = -3.381, p<0.001$).



\begin{figure}
    \centering
    \includegraphics[width=1\linewidth]{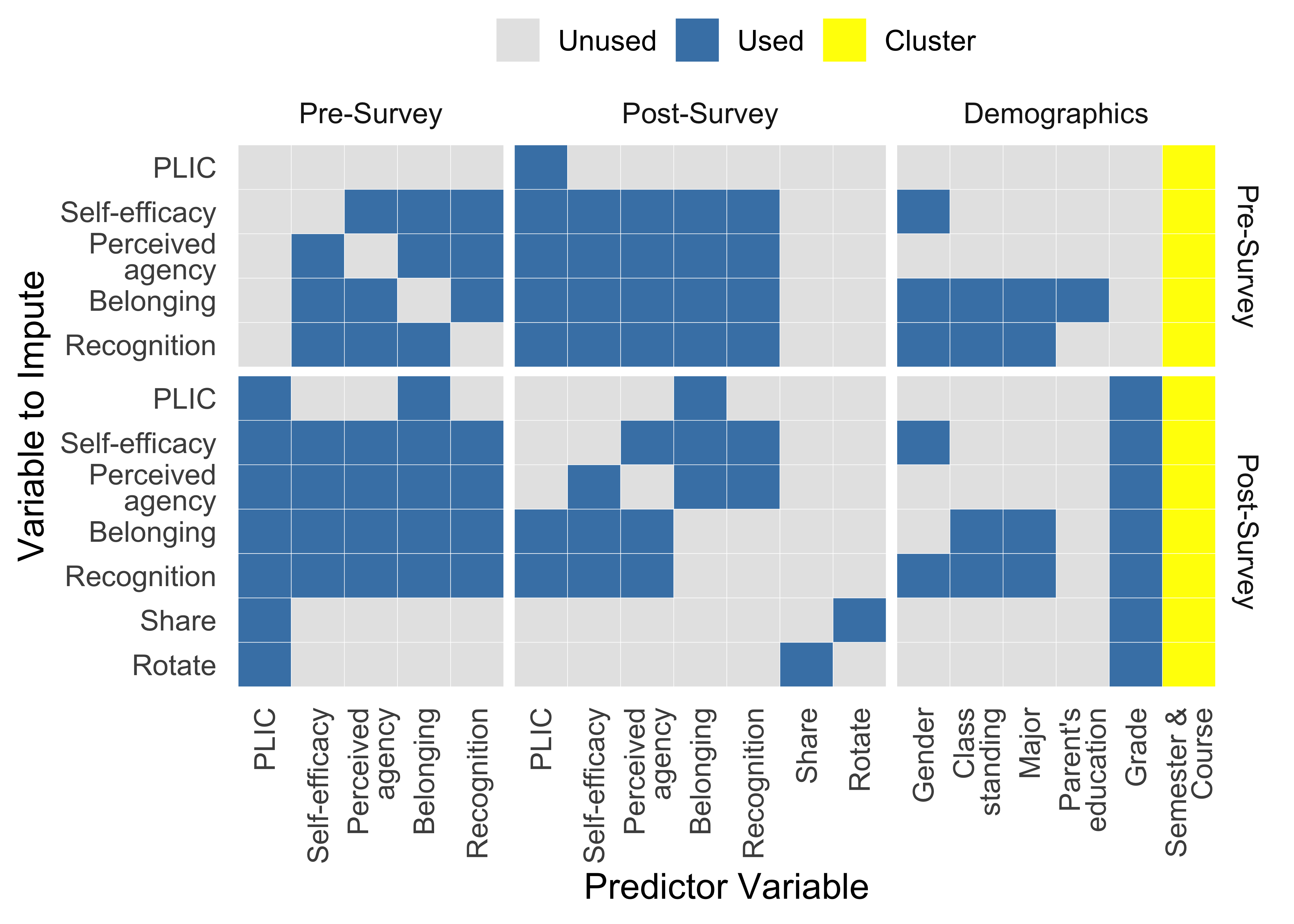}
    \caption{Predictor matrix for multiple imputation. ``Semester \& Course'' was used as a cluster variable for all variables. We passively imputed data for ``Split'' (not shown here).}
    \label{fig:predmatrix}
\end{figure}

These checks indicate variables that act as significant predictors for missingness across our data set. To prepare for imputation, we standardized all our variables to range from 0 to 1. Variables that were available for imputation are given in Table \ref{tab:imputation-variables}. We used the \verb|quickpred| function from the \textit{mice} package in $R$ to determine which variables should be used to impute for each variable \cite{mice_package}. Our predictor matrix is given in Figure \ref{fig:predmatrix}.

\begin{table}[]
\caption{Variables available for imputation for each course and semester. We only had access to student grades for course B.}
\label{tab:imputation-variables}
\begin{ruledtabular}
\begin{tabular}{lcccccc}
 & \multicolumn{3}{c}{Semester 1} & \multicolumn{3}{c}{Semester 2} \\ \cline{2-4} \cline{5-7}
Survey item & A1 & A2 & B & A1 & A2 & B\\ \hline
Pre-survey & & & & & & \\
\quad PLIC & X & X & X & X & X & X \\
\quad Self-efficacy & X & X & X & X & X & X \\
\quad Perceived agency & X & X & X & X & X & X \\
\quad Belonging & X & X & X & X & X & X \\
\quad Recognition & X & X & X & X & X & X \\
Post-survey & & & & & & \\
\quad PLIC  & X & X & X & X & X & X \\
\quad Self-efficacy & X & X & X & X & X & X \\
\quad Perceived agency & X & X & X & X & X & X \\
\quad Belonging & X & X & X & X & X & X \\
\quad Recognition & X & X & X & X & X & X \\
\quad Share & X & X & X & X & X & X \\
\quad Rotate & X & X & X & X & X & X \\
\quad Split &  &  &  &  &  &  \\
Demographics & & & & & & \\
\quad Gender & X & X & X & X & X & X \\
\quad Race/ethnicity & X & X & X & X & X & X \\
\quad Class standing & X & X & X & X & X & X \\
\quad Major & X & X & X & X & X & X \\
\quad Parent's highest level & X & X & X & X & X & X \\
\quad of education & & & & & & \\
\quad Grade & & & X & & & X \\
\end{tabular}
\end{ruledtabular}
\end{table}

We imputed all the pre- and post-semester outcome constructs (PLIC, self-efficacy, perceived agency, belonging, and recognition) as well as responses for the role distribution item. We used the imputation method \verb|2l.pmm| for the outcome constructs, which is predictive mean matching for multilevel data. Predictive mean matching uses values that exist within the data set, so it is more robust in cases of non-linear data and therefore more widely applicable \cite{vanginkel_2020}. 

Because the three group work role distribution choices (share, rotate, and split) must add up to 100\%, we used predictive mean matching imputation for the share and rotate items and then used passive imputation for the split item. Passive imputation simply means we imputed one value based on other values with a predefined equation. In this case, we defined Split as $1-\text{Share}-\text{Rotate}$.\footnote{We also technically imputed Grade, but this is due to requirements by the MICE package. To use Grade for Course B, MICE needs an imputed variable for Courses A1 and A2 grades. We do not discuss imputation for this variable since it acts as a dummy variable for these courses.}

According to prior recommendations \cite{madleydowd_2019, white_2011}, our number of imputations $m$ should meet the following threshold,
$$ FMI/m \leq 0.01 $$
where FMI is the Fraction of Missing Information \cite{madleydowd_2019, white_2011}. We tried 5, 10, 30, and 50 imputations and found that 50 was sufficiently many imputations so that every imputed variable had a low enough $FMI/m$ ratio. Our results use multiple imputation with $m=50$. Table \ref{tab:descriptive-statistics} shows the descriptive statistics across all courses, semesters, and constructs before and after imputation.

\begin{table*}[]
\caption{Descriptive statistics (mean and standard deviation) using listwise deletion and using multiple imputation for each course and semester.}
\label{tab:descriptive-statistics}
\begin{ruledtabular}
\begin{tabular}{lcccccc}
 & \multicolumn{3}{c}{Semester 1} & \multicolumn{3}{c}{Semester 2} \\ \cline{2-4} \cline{5-7}
Survey item & A1 & A2 & B & A1 & A2 & B \\
\hline
Using listwise deletion &  &  &  &  &  &  \\
Pre-survey &  &  &  &  &  &  \\
\quad PLIC & $0.39 \pm 0.12$ & $0.44 \pm 0.13$ & $0.42 \pm 0.13$ & $0.40 \pm 0.12$ & $0.45 \pm 0.12$ & $0.45 \pm 0.12$ \\
\quad Self-efficacy & $0.66 \pm 0.18$ & $0.73 \pm 0.18$ & $0.70 \pm 0.18$ & $0.68 \pm 0.18$ & $0.72 \pm 0.17$ & $0.70 \pm 0.15$ \\
\quad Perceived agency & $0.71 \pm 0.18$ & $0.72 \pm 0.20$ & $0.67 \pm 0.19$ & $0.71 \pm 0.19$ & $0.71 \pm 0.20$ & $0.69 \pm 0.18$ \\
\quad Belonging & $0.46 \pm 0.21$ & $0.59 \pm 0.23$ & $0.57 \pm 0.22$ & $0.46 \pm 0.21$ & $0.51 \pm 0.24$ & $0.56 \pm 0.20$ \\
\quad Recognition & $0.47 \pm 0.17$ & $0.47 \pm 0.21$ & $0.51 \pm 0.19$ & $0.47 \pm 0.19$ & $0.48 \pm 0.20$ & $0.54 \pm 0.18$ \\
Post-survey &  &  &  &  &  &  \\
\quad PLIC & $0.45 \pm 0.13$ & $0.45 \pm 0.13$ & $0.44 \pm 0.13$ & $0.43 \pm 0.13$ & $0.46 \pm 0.13$ & $0.52 \pm 0.14$ \\
\quad Self-efficacy & $0.82 \pm 0.14$ & $0.82 \pm 0.16$ & $0.84 \pm 0.14$ & $0.81 \pm 0.17$ & $0.80 \pm 0.16$ & $0.81 \pm 0.14$ \\
\quad Perceived agency & $0.78 \pm 0.19$ & $0.74 \pm 0.22$ & $0.76 \pm 0.19$ & $0.74 \pm 0.21$ & $0.74 \pm 0.22$ & $0.74 \pm 0.21$ \\
\quad Belonging & $0.59 \pm 0.26$ & $0.63 \pm 0.25$ & $0.67 \pm 0.22$ & $0.59 \pm 0.24$ & $0.60 \pm 0.24$ & $0.64 \pm 0.23$ \\
\quad Recognition & $0.53 \pm 0.21$ & $0.52 \pm 0.23$ & $0.56 \pm 0.21$ & $0.51 \pm 0.22$ & $0.51 \pm 0.23$ & $0.56 \pm 0.21$ \\
\quad Share & $0.49 \pm 0.31$ & $0.46 \pm 0.29$ & $0.42 \pm 0.25$ & $0.39 \pm 0.27$ & $0.39 \pm 0.30$ & $0.33 \pm 0.21$ \\
\quad Rotate & $0.21 \pm 0.22$ & $0.16 \pm 0.19$ & $0.26 \pm 0.19$ & $0.20 \pm 0.20$ & $0.17 \pm 0.21$ & $0.22 \pm 0.21$ \\
\quad Split & $0.31 \pm 0.29$ & $0.38 \pm 0.30$ & $0.32 \pm 0.25$ & $0.41 \pm 0.29$ & $0.43 \pm 0.32$ & $0.45 \pm 0.27$ \\
Using multiple imputation &  &  &  &  &  &  \\
Pre-survey &  &  &  &  &  &  \\
\quad PLIC & $0.38 \pm 0.13$ & $0.42 \pm 0.13$ & $0.41 \pm 0.13$ & $0.39 \pm 0.12$ & $0.42 \pm 0.13$ & $0.44 \pm 0.13$ \\
\quad Self-efficacy & $0.67 \pm 0.18$ & $0.72 \pm 0.18$ & $0.69 \pm 0.18$ & $0.68 \pm 0.18$ & $0.72 \pm 0.18$ & $0.71 \pm 0.16$ \\
\quad Perceived agency & $0.71 \pm 0.19$ & $0.72 \pm 0.20$ & $0.67 \pm 0.20$ & $0.71 \pm 0.20$ & $0.71 \pm 0.19$ & $0.70 \pm 0.18$ \\
\quad Belonging & $0.46 \pm 0.22$ & $0.55 \pm 0.24$ & $0.56 \pm 0.22$ & $0.46 \pm 0.21$ & $0.51 \pm 0.23$ & $0.56 \pm 0.20$ \\
\quad Recognition & $0.48 \pm 0.18$ & $0.49 \pm 0.21$ & $0.51 \pm 0.20$ & $0.47 \pm 0.19$ & $0.49 \pm 0.20$ & $0.54 \pm 0.18$ \\
Post-survey &  &  &  &  &  &  \\
\quad PLIC & $0.43 \pm 0.13$ & $0.44 \pm 0.13$ & $0.44 \pm 0.13$ & $0.43 \pm 0.13$ & $0.46 \pm 0.13$ & $0.50 \pm 0.14$ \\
\quad Self-efficacy & $0.81 \pm 0.16$ & $0.82 \pm 0.16$ & $0.83 \pm 0.15$ & $0.80 \pm 0.17$ & $0.82 \pm 0.16$ & $0.82 \pm 0.14$ \\
\quad Perceived agency & $0.77 \pm 0.20$ & $0.75 \pm 0.22$ & $0.76 \pm 0.20$ & $0.74 \pm 0.21$ & $0.75 \pm 0.22$ & $0.75 \pm 0.21$ \\
\quad Belonging & $0.58 \pm 0.25$ & $0.61 \pm 0.25$ & $0.66 \pm 0.23$ & $0.57 \pm 0.25$ & $0.60 \pm 0.25$ & $0.64 \pm 0.23$ \\
\quad Recognition & $0.52 \pm 0.22$ & $0.53 \pm 0.23$ & $0.56 \pm 0.22$ & $0.51 \pm 0.23$ & $0.52 \pm 0.23$ & $0.57 \pm 0.22$ \\
\quad Share & $0.48 \pm 0.30$ & $0.45 \pm 0.30$ & $0.43 \pm 0.27$ & $0.38 \pm 0.26$ & $0.37 \pm 0.26$ & $0.34 \pm 0.22$ \\
\quad Rotate & $0.19 \pm 0.20$ & $0.19 \pm 0.21$ & $0.25 \pm 0.20$ & $0.21 \pm 0.20$ & $0.20 \pm 0.21$ & $0.22 \pm 0.21$ \\
\quad Split & $0.33 \pm 0.28$ & $0.36 \pm 0.33$ & $0.32 \pm 0.26$ & $0.41 \pm 0.28$ & $0.43 \pm 0.30$ & $0.44 \pm 0.27$
\end{tabular} 
\end{ruledtabular}
\end{table*}

\subsection{\label{subsec:HLMAnalysis}HLM Analysis}
We evaluated the relationships between outcome constructs and group work role distributions using  hierarchical linear modeling~\cite{van_dusen_modernizing_2019, theobald_students_2018}. The hierarchical levels seek to account for potential systematic differences between the courses. These differences include the semester the course was offered, which may indicate whether students are in their first or second semester of college; the institution, with different institutions having different populations of students; and the course itself, as differences in the specific content covered in each course potentially leading to variations in student performance.

We used a two level model to connect a student's post-semester score on an outcome construct to their method of role distribution. We normalized the scores of the outcome  constructs (i.e. to a 0 to 1 scale) to account for differing scales between the PLIC and attitude constructs. We clustered responses based on the unique combination of course and semester. We modeled our level 1 data as in Equation \ref{eq:HLMLevelOne} -- the score for a post-semester construct, $Post$, for student $i$ in (semester + course) $j$ -- such that
\begin{align} \label{eq:HLMLevelOne}
    Post_{ij} &= \beta_{0j} + \beta_{1j} Pre + \beta_{2j} RD + r_{ij}
\end{align}
where $\beta_{0j}$ is the intercept; $Pre$ is the $i$th student's score for the pre-semester construct in (semester + course) $j$; $RD$ was the $i$th student's reported fraction of time spent on a given role distribution type in (semester + course) $j$; and $r_{ij}$ is the residual term.

We modeled our level 2 data as in Equation \ref{eq:HLMLevelTwo} such that,
\begin{align}  \label{eq:HLMLevelTwo}
    \beta_{0j} &= \gamma_{00} + u_{0j}
\end{align}
where $\gamma_{00}$ is the mean intercept and $u_{0j}$ is the residual term. We check that the assumptions of hierarchical linear modeling are met in Appendix \ref{sec:HLMAppendix}.

We theorized, based on previous work~\cite{doucette_share_2022}, that gender may be a significant factor that could also impact our model. Adding a gender term to the model, however, increased both AIC and BIC for all combinations of role distribution type and outcome construct for the data before multiple imputation. For the imputed data set, the AIC and BIC either increased or stayed approximately the same for all combinations of role distribution type and outcome construct. This indicates that the gender term did not improve the model. Thus, we did not include gender in our final model.

Beyond gender, we found insufficient theoretical motivation to include additional demographic characteristics (such as student race/ethnicity or major) as possible predictors of course outcomes. One study evaluated effects of student major (through lecture tracks) and found no consistent trend on lab group work preferences nor equipment handling~\cite{dew_group_2024}. Another study evaluated effects of race/ethnicity and found no effects on equipment handling~\cite{dew_so_2022}. We, therefore, excluded these other demographic variables from our models to avoid gap gazing~\cite{Traxler2016} and overfitting.

\section{\label{sec:Results}Results}
\subsection{Role Distributions}
\begin{figure*}
    \centering
    \includegraphics[width=1\linewidth]{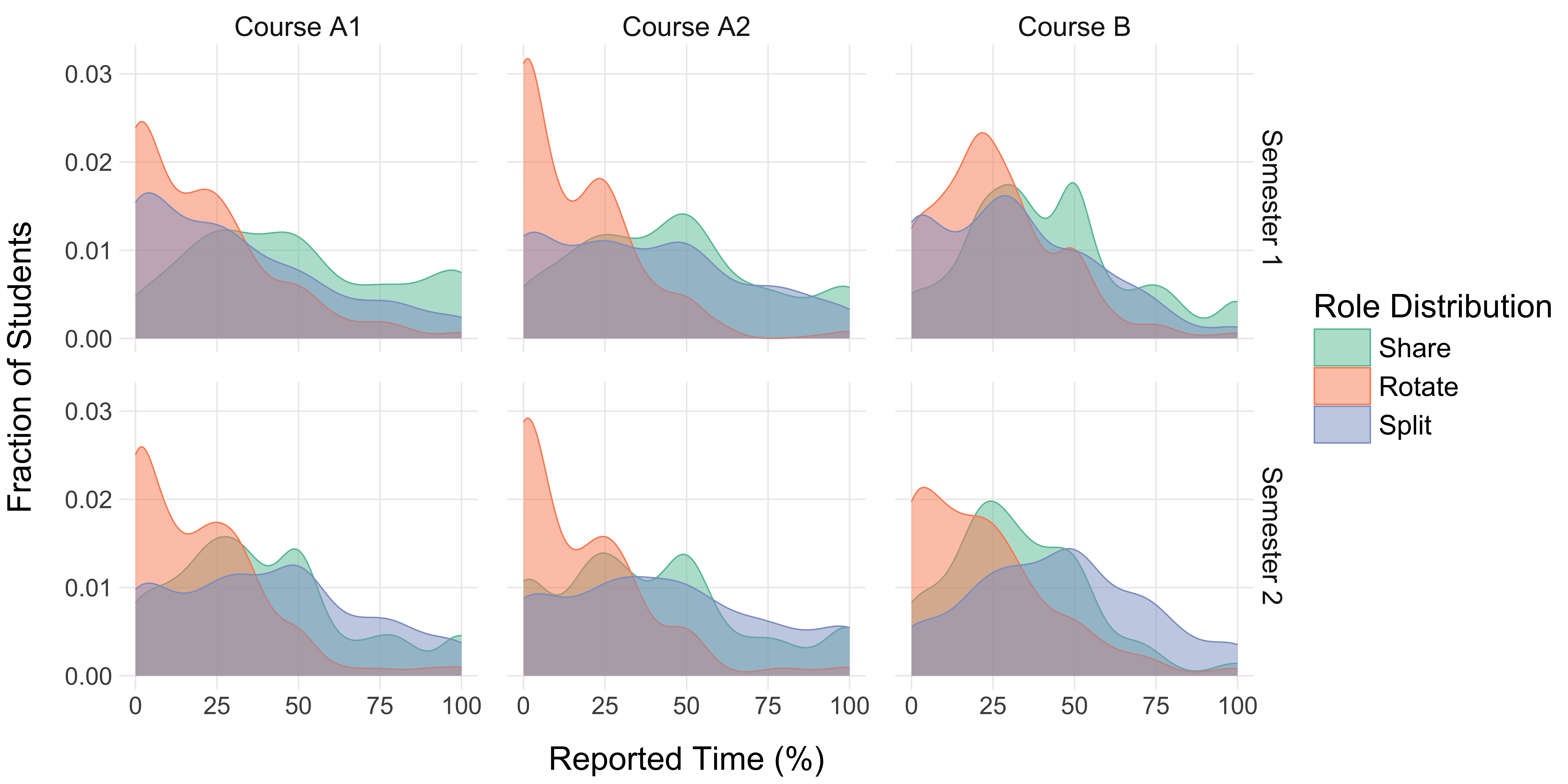} 
    \caption{Density plots of the reported percent of lab time students spent sharing, rotating, or splitting roles. Data presented are using listwise deletion.}
    \label{fig:roledist}
\end{figure*}

The results from the role distribution survey question before multiple imputation are shown in Fig. \ref{fig:roledist}. Across all courses in Semester 1, we found that students reported sharing roles (green) at higher rates than other distributions (evidenced by 
higher frequencies of greater reported times), followed by splitting (blue). In Semester 2, splitting was as frequent or more frequent than sharing. In both semesters, rotating (orange) was the least frequently reported role distribution (evidenced by higher peaks at the lowest reported times).




We found that across all classes, many students reported sharing roles between 25\% and 50\% of their time in lab. More students in Courses A1 and A2 than in Course B reported sharing roles for more than 50\% of the time. Course B had especially few students reporting sharing roles more than 50\% of the time in Semester 2 compared to any other course.

We found that many students reported rotating roles very rarely. In Courses A1 and A2 and Course B in Semester 2, the rotating curve peaks at the lowest reported times 
and very few students reported rotating roles for the majority of their time in lab. In Course B in Semester 1, however, the rotating curve peaks at about 25\% of the reported time.

Our data show that the percent of time in lab that students split roles was more evenly distributed compared to other role distribution methods. In Semester 1, we found that a greater fraction of students spent less than half of their time in lab splitting roles, whereas in Semester 2 we found that most students reported spending about half of their time in lab splitting roles.

\subsection{Construct Results}
\begin{figure*}
    \centering
    \includegraphics[width=1\linewidth]{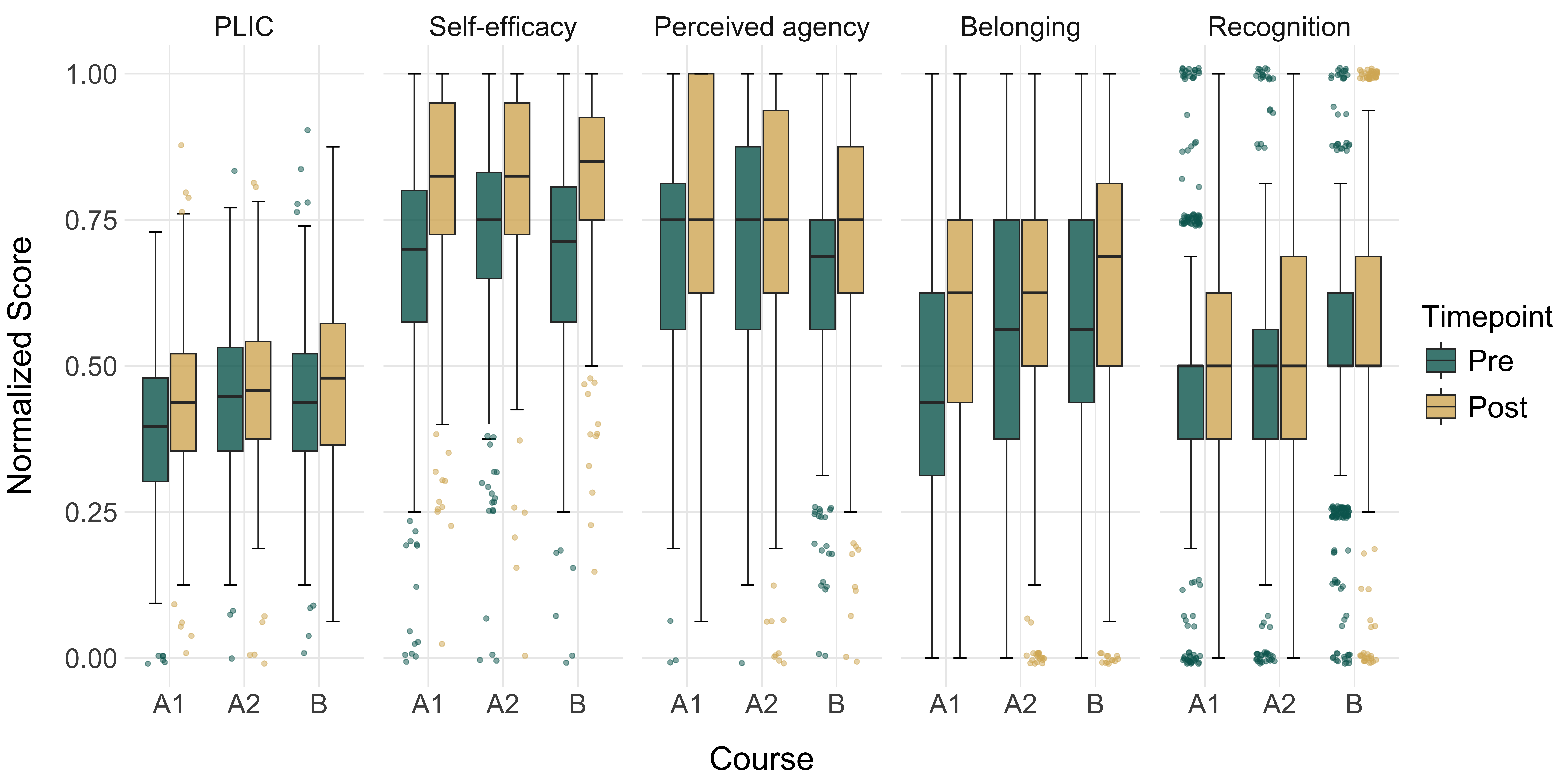}
    \caption{Bar graphs of the distribution of student scores on the PLIC and attitude constructs. The colored region represents the inter-quartile range and the horizontal line represents the median score. Whiskers indicate the values at $1.5$ times the interquartile range above the third quartile and below the first quartile. Dots indicate outliers defined as scores beyond the range of the whiskers (data presented is using listwise deletion).}
    \label{fig:constructs}
\end{figure*}

Students' overall pre- and post-survey scores on the outcome constructs are shown in Fig. \ref{fig:constructs}. We collapsed the data for Semesters 1 and 2 for each course because the score distributions were similar. We found that students' scores increase from pre- to post-survey for all constructs, though the size of the increases varied by course and construct. 
The attitudinal constructs tended to have skewed distributions with ceiling effects, particularly self-efficacy and perceived agency. 






\subsection{HLM Results}

\begin{table*}[]
\begin{ruledtabular}
\caption{Results from hierarchical linear modeling for change in outcome construct scores. The table includes the standardized effect sizes, standard errors, and $p$ values in parentheses.}
\label{tab:HLMresult}
\begin{tabular}{lccc}
Construct & Share & Rotate & Split\\ \hline
Using listwise deletion & & & \\
\quad PLIC & 0.0579 $\pm$ 0.0766 (0.450) & -0.2321 $\pm$ 0.1029 (0.024) & 0.0645 $\pm$ 0.0733 (0.379)\\
\quad Self-efficacy & 0.2443 $\pm$ 0.0725 (0.001) & 0.2276 $\pm$ 0.0979 (0.020) & -0.3396 $\pm$ 0.0690 ($<0.001$)\\
\quad Perceived agency & 0.1695 $\pm$ 0.0735 (0.021) & 0.2286 $\pm$ 0.0997 (0.022) & -0.2725 $\pm$ 0.0702 ($<0.001$)\\
\quad Belonging & 0.2336 $\pm$ 0.0689 (0.001) & -0.1226 $\pm$ 0.0933 (0.189) & -0.1532 $\pm$ 0.0661 (0.021)\\
\quad Recognition & 0.1779 $\pm$ 0.0700 (0.011) & 0.1783 $\pm$ 0.0949 (0.060) & -0.2557 $\pm$ 0.0668 ($<0.001$)\\
Using multiple imputation & & & \\
\quad PLIC & 0.0041 $\pm$ 0.0642 (0.949) & -0.1237 $\pm$ 0.0923 (0.181) & 0.0574 $\pm$ 0.0650 (0.378)\\
\quad Self-efficacy & 0.1347 $\pm$ 0.0626 (0.032) & 0.0968 $\pm$ 0.0845 (0.253) & -0.1705 $\pm$ 0.0580 (0.003)\\
\quad Perceived agency & 0.1009 $\pm$ 0.0604 (0.095) & 0.0926 $\pm$ 0.0846 (0.274) & -0.1378 $\pm$ 0.0571 (0.016)\\
\quad Belonging & 0.1231 $\pm$ 0.0638 (0.054) & -0.0600 $\pm$ 0.0810 (0.459) & -0.0820 $\pm$ 0.0580 (0.158)\\
\quad Recognition & 0.1096 $\pm$ 0.0593 (0.065) & 0.0581 $\pm$ 0.0810 (0.474) & -0.1281 $\pm$ 0.0583 (0.028)\\
\end{tabular} 
\end{ruledtabular}
\end{table*}

\begin{figure*}
    \centering
    \includegraphics[width=1\linewidth]{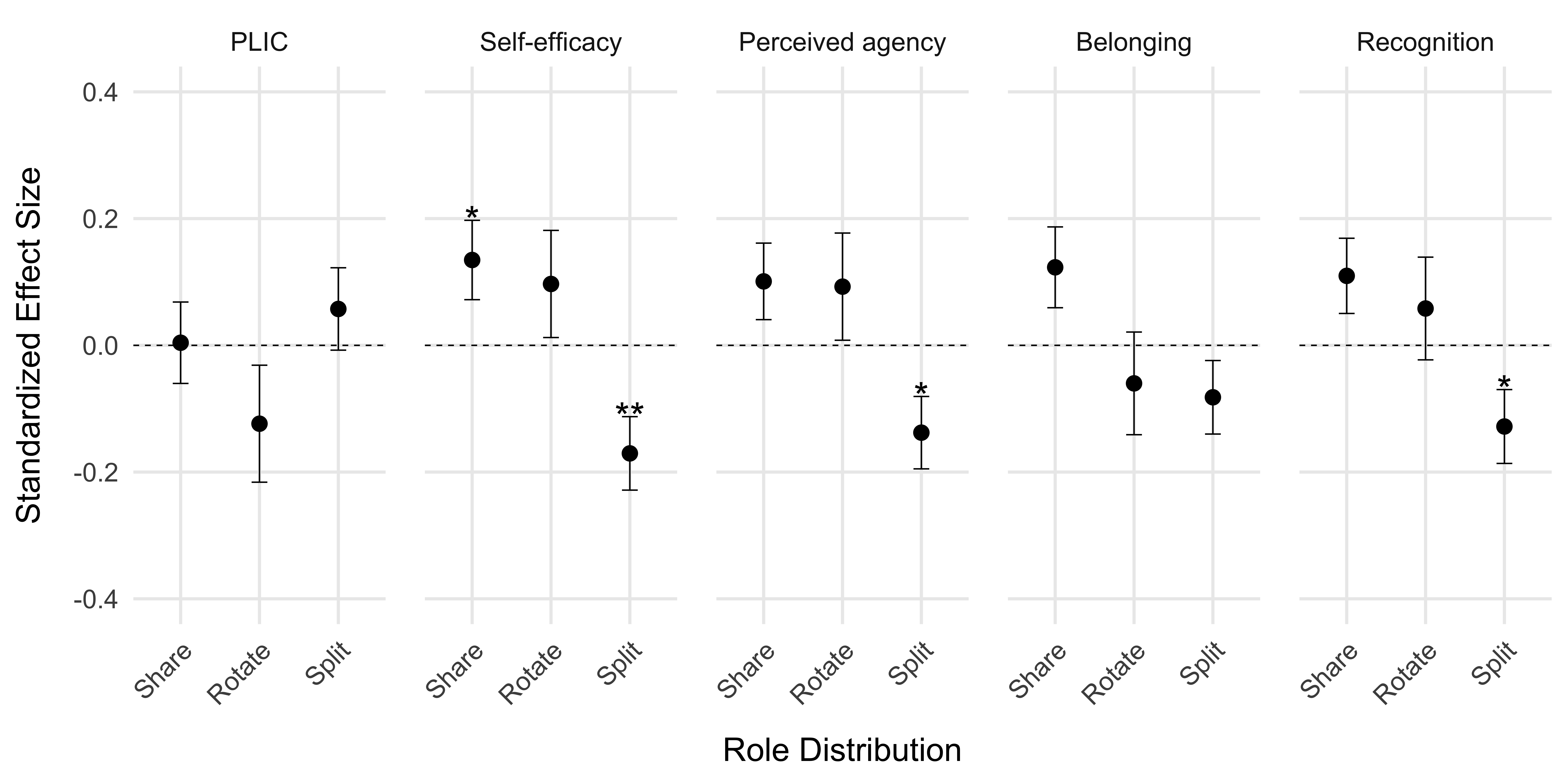}
    \caption{Results of hierarchical linear modeling of change in PLIC and attitude scores controlling for semester and course using data after multiple imputation. A value greater than zero indicates a positive effect on score, whereas a value less than zero indicates a negative effect on score. The horizontal dashed line indicates no observable effect on score. The error bars represent the standard error of the effect size. The asterisks denote statistical significance where $*$ indicates $p < 0.05$, and $**$ indicates $p < 0.01$.}
    \label{fig:hlm_results}
\end{figure*}

The results of the HLM analysis are in Figure \ref{fig:hlm_results} and Table \ref{tab:HLMresult}. The figure indicates the effect size for each outcome construct, such that a positive effect means more reported time in a particular role correlates to higher post-test scores, controlling for pre-test scores and random effects. We found that a student's role distribution had no significant effect on their PLIC performance. 
Student role distributions did have an effect on attitude constructs, with similar trends across each construct, but varying statistical significance. In all cases, more time spent sharing roles corresponded to the strongest positive effect sizes (statistically significantly greater than zero only for self-efficacy). More time spent splitting roles corresponded to the strongest negative effect sizes (statistically significantly less than zero for self-efficacy, perceived agency, and recognition). The effect of rotating varied between constructs (sometimes positive and sometimes negative effect sizes), but was never statistically distinguishable from zero.

\section{\label{sec:Discussion}Discussion}

In this study, we evaluated the effect of group work role distributions (whether sharing, rotating, or splitting hands-on lab roles) on several student outcomes: critical thinking, self-efficacy, perceived agency, belonging, and recognition. Here we analyze our results through the lens of our research questions.

\subsection{Group Roles Distributions}

Despite different student populations, we found that students across courses and semesters reported sharing or splitting roles much more frequently than they rotated roles. 
Students' role distributions may relate to their group work preferences. At Institution A, a previous study \cite{dew_group_2024} found that students preferred sharing and splitting roles, with the fewest students preferring to rotate roles, consistent with their reported role distributions here. At Institution B, however, a previous study \cite{holmes_evaluating_2022} found that most students---especially women---preferred sharing roles. These preferences are consistent with the data from Semester 1, but not from Semester 2.

Alternatively, prior research suggests that student preferences do not sufficiently predict the roles that they actually take on~\cite{holmes_evaluating_2022, dew_group_2024}. An alternative explanation of the role distributions data, therefore, is the use of partner agreements~\cite{dew_group_2024}. Courses A1 and A2 as well as Semester 2 of Course B all used partner agreements and their distributions look similar in Fig. \ref{fig:roledist}. Semester 1 of Course B, however, did not use partner agreements and looks slightly different than the other courses; particularly more students reported rotating roles. While we do not have data on student preferences for these semesters, we can reasonably infer that preferences are similar across semesters within each course.
This suggests partner agreements may have a larger effect on student role distributions. Future work should test this hypothesis directly. 


We also note spikes in the frequency of students who reported rotating roles 25\% of the time in lab, splitting between 25\% and 50\% of the time, and sharing 50\% of the time. Despite our survey question offering students a continuous range of reported times, it is possible students divided their reported time into discrete chunks, such as quarters, halves, or thirds, resulting in these spikes. It is also possible many 
students were reluctant to select the most extreme values for the percentage of time spent using each type of role distribution, \textit{i.e.,} 0\% and 100\%. Thus, analyzing the data along a continuous scale could overestimate the precision of students' perceptions of how they spent their time. The distributions themselves, however, suggest a reasonably continuous distribution of student responses that are sufficient for our linear regression analyses. Thus, we do not believe this limitation affected our analysis, particularly since our models pass our assumption tests (see Appendix \ref{sec:HLMAppendix} and the Supplemental Material \cite{supplementalMaterial}).


\subsection{Outcome Constructs}

We find that student scores increased from pre- to post-test across all measures and in all courses. The current study is the first study to report data about students' responses to the lab-specific self-efficacy, belonging, and recognition items, though we can compare the shifts in PLIC and perceived agency scores to previous studies using these same constructs.

The raw gains in PLIC scores ranged from 0.01 to 0.07 (Table~\ref{tab:descriptive-statistics}), as compared to the average raw gains of 0.08 across various types of labs reported in Ref.~\cite{walsh_quantifying_2019}\footnote{Note: Two items were removed from the PLIC after Ref.~\cite{walsh_quantifying_2019} was published based on work in Ref.~\cite{heim_what_2022}. The average raw gain of 0.08 was estimated from the reported item scores, excluding the two removed items.}. The pre-survey scores in the current study were on par with those reported in Ref.~\cite{walsh_quantifying_2019}, but the post-survey scores were lower, suggesting instruction in these courses was on the lower end of effectiveness at shifting students' critical thinking scores. 

The raw gains in the perceived agency items ranged from 0.02 to 0.09 (out of 1; Table~\ref{tab:descriptive-statistics}), as compared to the average raw gains of 0.2 reported in Ref.~\cite{kalender_restructuring_2021}. The pre-survey perceived agency scores in our study were higher (around 0.7) than those reported in Ref.~\cite{kalender_restructuring_2021} (which were around 0.5). Our post-survey scores were similar to theirs (around 0.75 and 0.70, respectively). This suggests instruction in both studies reached similar levels of perceived agency, or that the survey items reached a ceiling. 

We do not see large differences in score increases between courses, which can likely be attributed to the courses using a similar pedagogy~\cite{Smith2020}. We see several limitations towards the interpretability of these results, however. First, the self-efficacy and perceived agency scores show clear ceiling effects. These measures may need to be recalibrated to have a more normal distribution centered around 0.5. Future work should evaluate these measures in other pedagogical and instructional contexts to evaluate the generalizability of this ceiling effect as well as to evaluate the range of student gains across instructional lab pedagogy. 

\subsection{Group Role Effects}

Our main result is the impact of the group work role distributions on student outcomes. First, we find no effect of role distribution on PLIC outcomes. Second, we find a consistent trend (with varied significance) between role distributions on the attitudes constructs: 
sharing roles has the strongest positive effect, splitting has the strongest negative effect, and the effect of rotating lies in between.

For PLIC scores, based on the ICAP framework, we expected sharing roles to have a more positive effect than rotating, interpreting shared roles as better for fostering constructive dialogue than the other roles. We also expected splitting roles to have the smallest impact, as the division of roles would support individual experts and not necessitate constructive dialogue. Compared to splitting roles, rotating roles means all students develop expertise across roles, creating opportunities for constructive dialogue between students. Our results, however, show no significant differences in the effects of role distribution on PLIC scores. 
This null effect suggests that group work role distributions do not differently develop critical thinking as measured by the PLIC. We consider two potential explanations. The first relates to measurement precision. The gains in PLIC scores of this sample of students may be too small to allow us to detect effects of role distribution; the mean PLIC scores only increase from 0.01 to 0.07 (out of 1) across courses. The second potential explanation is that groups' \textit{hands-on} role distributions do not affect critical thinking. 
While students may have distributed the hands-on aspects of roles one way, they might have distributed the \textit{decision-making} aspects 
differently. For example, a group of students may have split into an equipment handler, a note-taker, and a data analyst for hands-on roles, but jointly decided how they would take measurements, write their results, and present their data. In this case, students would have limited hands-on access to roles, but still gain experience in the critical thinking associated with each role. In the ICAP framework, students collaboratively making decisions regarding different roles in the lab would be classified as interactive, the tier with the highest engagement. Thus, it is possible that the distribution of decision-making for roles has a greater impact on critical thinking than the distribution of hands-on roles, a distinction that would not be captured by our survey item.


For self-efficacy, we find that sharing roles had a positive, statistically significant effect and splitting roles had a negative, statistically significant effect. Rotating roles had a positive -- but not statistically significant -- effect. We expect that sharing roles had a net positive effect because each student spends their time on each role every lab session. When rotating roles, a student spends their time on one role each lab session. When splitting roles, students spend their time on only one role throughout the semester. We expect that students who spend more time in a role would have greater self-efficacy in that role. Many of the self-efficacy items in the survey ask about student confidence regarding specific tasks; for example, students are asked if they can ``comfortably take charge of the equipment'' and ``interpret graphs.'' On average, students in groups who share more frequently should feel confident with a larger variety of tasks. Conversely, on average, students in groups who only gain experience with one role, should feel less confident with tasks outside their role.


The trends in perceived agency are similar to those seen with self-efficacy; though none of the effects are statistically significant for perceived agency. The similar trends matches our expectations because the two attitudes are theoretically connected~\cite{bandura_self-efficacy_1982}. 

The shift in significance may have several explanations. First, most of the self-efficacy items probe students about specific tasks, while the perceived agency items are more broad. Plausibly, students are answering these items considering their perceived agency in the course overall, which would be similar across all role distributions. Any differences from the group work role distributions would need to be measured by role-specific items so that students evaluate their perceived agency within their lab group. 
Future work could verify the theory that the specificity of self-efficacy items leads to more pronounced gains compared to perceived agency by measuring the gain in each self-efficacy item against the amount of time a student spends in each role or by crafting role-specific perceived agency items that reflect the self-efficacy items.

For belonging, none of the group work role distributions had a statistically significant effect on belonging, though we again see that sharing had a positive effect and splitting had a negative effect. 
Research in cooperative learning~\cite{johnson_educational_2009} suggests that group processing~\cite{bertucci_influence_2012}, through targeted reflection activities, impacts students' sense of belonging~\cite{strahm_cooperative_2007}. We expect that sharing roles, or being in a group that shares roles, more likely creates opportunities for group processing than the other group work role distributions, explaining the more positive impact on belonging. We may not have observed a statistically significant effect because the labs did not implement a structure to ensure students cooperated and engaged in explicit group processing. 


For recognition, the statistically significant, negative effect of splitting on recognition could be explained by students who are siloed into specific roles not feeling like they are doing a ``physics'' role. This siloing may lead some students to feel that their peers and instructors do not see them as physics people. We would then expect recognition gains to vary based on a student’s role in their groups when they split roles. Based on the findings of \cite{doucette_hermione_2020, Pettersson_2011, danielsson_exploring_2012, gonsalves_masculinities_2016}, we expect that roles like note-taking would have lower recognition gains than roles like equipment. Future work could verify this theory by measuring recognition gains against the amount of time a student spends in each role. Alternatively, aspects of recognition may be associated with feelings of acceptance by their peers, which is connected to feelings of belonging in cooperative learning~\cite{strahm_cooperative_2007}. This explanation supports our findings above that the patterns for recognition are similar to those for belonging.

Altogether, these findings and their associated limitations motivate a range of future research. First, because our results are observational, our findings are inevitably intertwined with student choice and preference. Future studies should implement random assignment of lab groups to share, rotate, or split their lab roles throughout a semester to disentangle the isolated effect of group choice on outcome. Research should also expand to more lab environments with different group sizes, lab pedagogies, and student populations. Future work could also use observations of student work in lab instead of student self-reports. Importantly, these results support the findings of \cite{doucette_share_2022} with a larger data set and additional measurement constructs. 

\begin{acknowledgments}
We would like to thank Andrew Loveridge and Viranga Perera for assisting with data collection by distributing surveys. We would also like to thank Matt Thomas and the Cornell Statistical Consulting Unit for their assistance with data analysis. Finally, we would like to thank Z. Yasemin Kalender and Martin Stein for their work on survey development. This material is based upon work supported by the National Science Foundation Grant No. DGE-2139899 and the Cornell Nexus Scholars Program.
\end{acknowledgments}

\section*{Data Availability}
The data that support the findings of this article are openly available \cite{supplementalMaterial}.

\appendix
\section{Assumptions for Hierarchical Linear Modeling}\label{sec:HLMAppendix}
Here, we check assumptions required for hierarchical linear modeling following prior recommendations \cite{van_dusen_modernizing_2019}. In the main text, we analyze 15 different regression models: three role distributions for five outcome constructs. Here, we provide plots for only three role distributions for a single outcome construct: the PLIC. The other figures are available in the Supplemental Material \cite{supplementalMaterial}.

We analyze regression models for Equation \ref{eq:HLMLevelOne} in the case of sharing roles, as in Equation \ref{eq:HLM_PLIC_Share}.
\begin{align} \label{eq:HLM_PLIC_Share}
    \text{Post-PLIC}_{ij} &= \beta_{0j} + \beta_{1j} \text{Pre-PLIC} + \beta_{2j} \text{Share} + r_{ij}
\end{align}
Next, in the case of rotating roles, as in Equation \ref{eq:HLM_PLIC_Rotate},
\begin{align} \label{eq:HLM_PLIC_Rotate}
    \text{Post-PLIC}_{ij} &= \beta_{0j} + \beta_{1j} \text{Pre-PLIC} + \beta_{2j} \text{Rotate} + r_{ij}
\end{align}
And finally, in the case of splitting roles, as in Equation \ref{eq:HLM_PLIC_Split}.
\begin{align} \label{eq:HLM_PLIC_Split}
    \text{Post-PLIC}_{ij} &= \beta_{0j} + \beta_{1j} \text{Pre-PLIC} + \beta_{2j} \text{Split} + r_{ij}
\end{align}

We do not know of any recommendations regarding whether to check the assumptions of hierarchical linear modeling using data before or after multiple imputation. Following another study \cite{nissen_2021_investigating}, we check our assumptions using the averaged results from the pooled data set.

\begin{figure}
\includegraphics[width=\columnwidth]{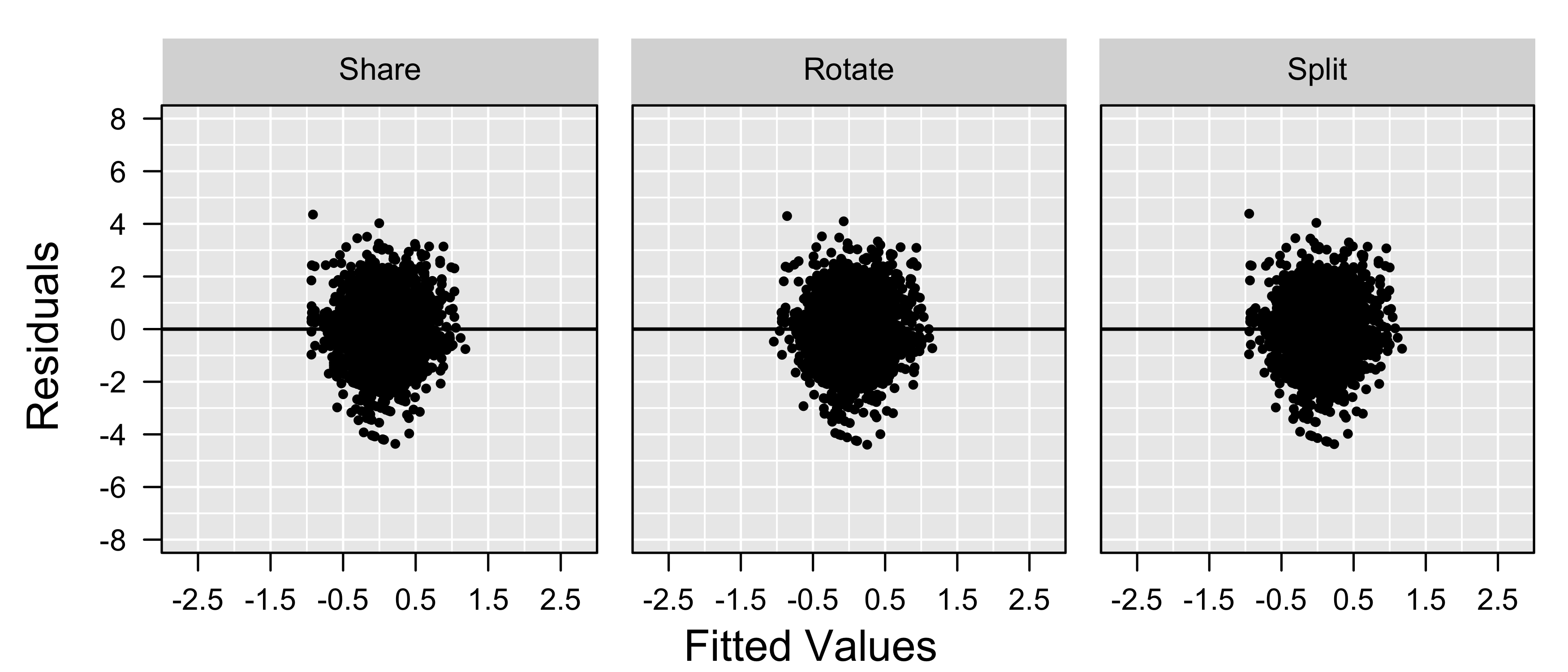}
\caption{\label{HLM_Linearity_PLIC}A visual check for hierarchical linear modeling's assumption of linearity for the PLIC models using data after multiple imputation. These plots display the residuals versus fitted values for three different models: share (see Equation \ref{eq:HLM_PLIC_Share}), rotate (see Equation \ref{eq:HLM_PLIC_Rotate}), and split (see Equation \ref{eq:HLM_PLIC_Split}).}
\end{figure}

We check the assumption of linearity in Figure \ref{HLM_Linearity_PLIC}. We see no trends in this data, suggesting it meets the assumption of normality. For the other constructs, we do find some ceiling and floor effects (as seen in the original data in Fig. \ref{fig:constructs}) but do not see any other trends in the plots.

For the assumption of homogeneity of variance, we obtained no statistically significant results from our ANOVA (share returned $p = 0.7399$, rotate returned $p=0.7399$, and split returned $p = 0.7387$). We found no statistically significant effects for any of the three role distribution models across the four attitude constructs ($p>=0.5674$) which means our data meet the assumption of homogeneity of variance in all cases. 

\begin{figure}
\includegraphics[width=\columnwidth]{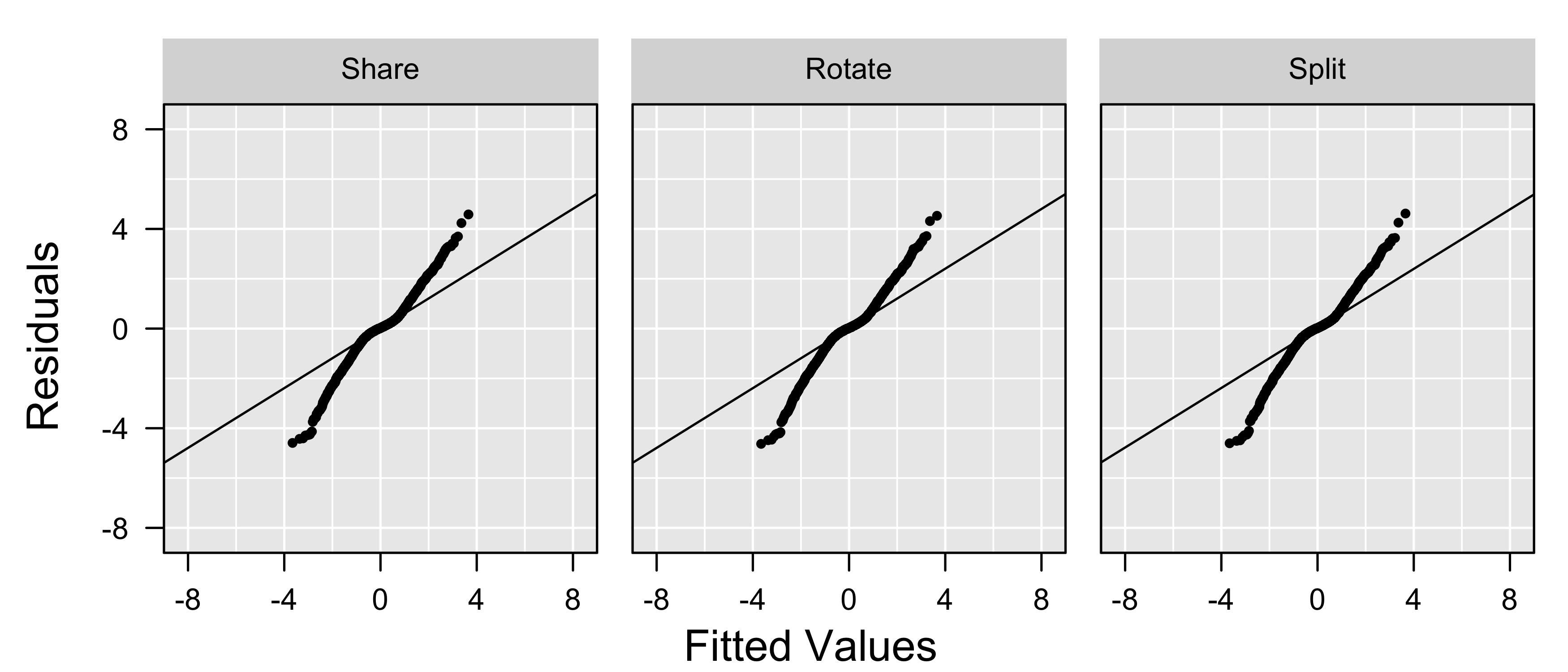}
\caption{\label{HLM_Normality_PLIC}A visual check for hierarchical linear modeling's assumption of normality for the the PLIC models using data after multiple imputation. These plots display the residuals versus fitted values for three different models: share (see Equation \ref{eq:HLM_PLIC_Share}), rotate (see Equation \ref{eq:HLM_PLIC_Rotate}), and split (see Equation \ref{eq:HLM_PLIC_Split}).}
\end{figure}

We check the assumption of normality in Figure \ref{HLM_Normality_PLIC}. While we see the residuals stray from normality at the ends of the distribution, this usually does not affect interpretations of significance \cite{Lumley_2002, walsh_skills-focused_2022} and has been seen in similar physics education research studies \cite{walsh_skills-focused_2022, dew_group_2024, dew_structuring_2025}. We see somewhat similar trends for our attitude constructs. 

We used the \verb|lmer| function from the \textit{lme4} package in R for our hierarchical linear models \cite{lme4Package}.

\bibliography{references.bib, apssamp.bib}

\end{document}